# The First Photometric and Spectroscopic Study of Contact Binary V2840 Cygni

Ravi Raja Pothuneni, Shanti Priya Devarapalli, and Rukmini Jagirdar

Department of Astronomy, Osmania University 500007, India; ravirajapothuneni@osmania.ac.in

## Abstract

The first photometric, spectroscopic and period variation studies of neglected short-period eclipsing binary V2840 Cygni are presented. High mass ratio contact binaries (HMRCBs), especially those in the weak-contact configuration, are vital when probing the evolutionary models of contact binaries (CBs) using stellar models. The photometric solutions reveal the weak-contact nature of V2840 Cygni with a high mass ratio (~1.36), motivating us to investigate the nature of such binaries. The period variation study of V2840 Cygni spanning 15 yr shows a secular period decrease at a rate of $\sim 5.5 \times 10^{-7}$ day $yr^{-1}$, indicating mass transfer between the components. The superimposed cyclic variation provides a basic understanding of the possible third body ($P_3 \sim 8$ yr, $m_3 \sim 0.51\ M_\odot$). Following the derived parameters, the evolution of the system is discussed based on the thermal relaxation oscillation (TRO) model. It is found that V2840 Cygni falls in a special category of HMRCBs, which validates TRO. To characterize the nature of HMRCBs, a catalog of 59 CBs with high mass ratios has been compiled along with their derived parameters from the literature. For all the HMRCBs in the study, a possible correlation between their contact configuration and observed period variations for relative log $J_{rel}$ is discussed. The spectroscopic study of V2840 Cygni provides evidence of the presence of magnetic activity in the system and the existence of ongoing mass transfer which is additionally deduced from the period variation study. The LAMOST spectra of 17 HMRCBs are collected to interpret the stellar magnetic activity in such systems.

## 1. Introduction

Contact binaries (CBs) of EW-type are significant in studying and understanding their fundamental stellar properties and evolution (Lucy & Wilson 1979; Priya et al. 2020) using high precision photometric and spectroscopic observations. They consist of two stars filling or overfilling their Roche lobes while having a common convective envelope around both the components, leading to mass and energy transfer between them. The angular momentum and mass transfer/loss are prominently defined for these systems as due to various evolutionary processes such as thermal relaxation oscillation (TRO) cycles (Flannery 1976; Lucy 1976; Robertson & Eggleton 1977) and angular momentum loss (AML) (Vilhu & Rahunen 1980; Qian 2003). The presence of a third body (Devarapalli et al. 2020) and the Applegate mechanism (Applegate 1992) are further analyzed through orbital period variation studies. They are rich sources not only for examining the stellar properties but also for testing the current theories on binary evolution, interaction between their components and the influence of additional component(s) (Zakirov 2010), if any.

CBs have been categorized on the basis of their mass ratios as low mass ratio (LMR) CBs, with $q \leqslant 0.4$ and high mass ratio

(HMR) CBs, with $q > 0.4$ (Qian 2001). While Rucinski (2001) observed that the fraction of high mass ratio contact binaries (HMRCBs) seems to be lower than that of low mass ratio contact binaries (LMRCBs), Webbink (1979), Mochnacki (1981), Csizmadia & Klagyivik (2004) emphasized that the class of HMRCBs with $q \geqslant 0.7$ behaves in high contrast to other CBs and exhibits very poor energy transfer rates. Their formation and evolution are still open questions. Based on the period variation studies, they are known to undergo evolution either through TRO or AML-controlled processes. As per the TRO-model (Flannery 1976; Lucy 1976; Robertson & Eggleton 1977), HMRCBs with a period increase are predicted to undergo TRO cycles around the marginal-contact phase and oscillate between the semi-detached and contact configurations (e.g., Liu et al. 2007, 2016; He & Qian 2009; Zhu et al. 2010, 2013b; Dai et al. 2019). Whereas, in the AML-model (Vilhu & Rahunen 1980; Qian 2003), LMRCBs with period decrease undergo AML via magnetic braking by stellar wind (e.g., Bradstreet & Guinan 1994; Stepien 2005; Yakut & Eggleton 2005; Sarotsakulchai et al. 2019; Devarapalli et al. 2020) and oscillate around a critical mass ratio evolving into a deeper configuration with an HMR and period decrease (Qian et al. 2013b). There are very few HMRCBs found undergoing TRO-controlled evolution and thus they become great targets for testing the TRO theory. Here, the mass exchange in the previous semi-detached phase from the hotter to the cooler component occurs very rapidly. As a result, the mass-accreting component loses thermal equilibrium, expands rapidly and then the system slowly comes into a contact phase exhibiting period decrease. The components then slowly traverse to a semi-detached configuration again with an increase in period. So, studying the HMRCBs, especially in the weak-contact configuration, becomes vital in understanding this phase in CB evolution (He & Qian 2009). In the current work, we have attempted to study one such star, V2840 Cygni (R.A. $21^h49^m13.^s21$, decl. $+30°58'05.''95$, TESS$_{Meanmag}$ = 11.63 and $T_h$ = 6657 K).

V2840 Cygni is one of the variables in our list of neglected objects being studied with the latest photometric data available from Transiting Exoplanet Survey Satellite (TESS) and Gaia Early Data Release 3 (EDR3) epoch photometry databases, planned for extensive investigations to understand their origins and evolution. V2840 Cygni is a short period eclipsing binary ($P \sim 0.6311$ day) cataloged for the first time as an EW-type variable star in the General Catalogue of Variable Stars (Kazarovets et al. 2017) with $V \sim 11.81$ (Heinze et al. 2018). Except for cataloging, no photometric or spectroscopic analysis is reported in literature. Hence, a detailed study was carried out using the available photometric data and ground-based spectroscopic observations. From the obtained results, the model explaining the evolution of V2840 Cygni is discussed in detail in the current work.

## 2. Data Collection and Reduction for V2840 Cygni

### 2.1. TESS and Gaia Epoch Photometry

The high precision 30 minute cadence data from TESS observed in Sector 15 during 2019 August 15–September 10 were used for the photometric analysis of V2840 Cygni. The light curve data were produced by the Quick-Look Pipeline (QLP) (Huang et al. 2020a, 2020b; Kunimoto et al. 2021), made publicly available and accessible through the Mikulski Archive for Space Telescopes (MAST) portal.[1] Kepler Science Pipeline Simple Aperture Photometry Flux (KSPSAP_FLUX), which is a normalized light curve detrended by kepler spline (Huang et al. 2020b), was used for the light curve modeling. A total of 735 data points covering an observation period of about 22 days were collected for the light curve analysis.

The Gaia EDR3 (Brown et al. 2021) contains the data accumulated during the first 34 months of the nominal mission (The Gaia mission, Prusti et al. 2016). The epoch photometric data are provided as multi-color photometric time series in the broad optical photometric bands of $G$, and the blue $G_{BP}$ and red $G_{RP}$ bands. For the light curve analysis of V2840 Cygni, the red $G_{RP}$ band photometric data, processed by the Gaia Data

Processing and Analysis Consortium (DPAC) (Siopis et al. 2020), were used. A total of 63 data points observed between 2014 and 2017 were collected for the analysis. The detailed light curve analysis of V2840 Cygni is given in Section 3.1.

### 2.2. Times of Minima

A total of 39 times of primary minima (hereafter ToMs) were extracted from the latest TESS, ASAS-SN and SuperWASP data, spanning a period of 15 yr as listed in Table 1. The ToMs were determined following a standard procedure (Hoffman et al. 2006; Papageorgiou et al. 2021) of fitting a Gaussian function to the primary eclipses in the observed data. The period ($P$) of the system was evaluated and determined using Period04 (Lenz & Breger 2005). The least-squares fitting method was utilized to determine the ephemeris (BJD$_0$) of V2840 Cygni from TESS data and is given by

$$\text{Min.I} = \text{BJD}2458731.^d780456(\pm0.020833) \\ + 0.^d631123(\pm0.004290) \times E, \quad (1)$$

where $E$ is the Epoch number.

The above data (Table 1) were used for period variation study and are discussed in Section 6.

### 2.3. Spectroscopy

The spectroscopic observations of V2840 Cygni were taken using the medium resolution Hanle Faint Object Spectrograph Camera instrument equipped with a 2k × 4k CCD detector on the 2 m Himalayan Chandra Telescope (HCT) at Indian Astronomical Observatory (IAO), Hanle, India on 2021 September 30. Five spectra of V2840 Cygni along with a spectroscopic standard (HD 203454, R.A. $21^h21^m01.^s41$, decl. $+40°20'41.''90$) were obtained using the Gr7 grism in the spectral range of 380–684 nm at a resolution of ∼1300. Zero exposure frames, halogen lamp spectra and FeAr lamp spectra were obtained for bias subtraction, flat-fielding and wavelength calibration respectively.

IRAF[2] package "*spectred*" was employed for the reduction and calibration of spectroscopic data which were then normalized for further studies. A sample spectrum showing characteristic lines of V2840 Cygni in comparison to its spectral standard star (HD 203454, R.A. $21^h21^m01.^s41$, decl. $+40°20'41.''90$), was used for spectral line analysis.

## 3. Data Analysis of V2840 Cygni

### 3.1. Light Curve Analysis

Due to the unavailability of radial velocity data, light curve analysis of V2840 Cygni was performed on the latest available

---

[1] https://mast.stsci.edu/portal/Mashup/Clients/Mast/Portal.html

**Table 1**
The Primary Times of Minima for V2840 Cygni

| BJD$_0$ (2,450,000+ days) | Error | $E$ | $(O-C)$ (days) | Res (days) | Survey | BJD$_0$ (2,450,000+ days) | Error | $E$ | $(O-C)$ (days) | Res (days) | Survey |
|---|---|---|---|---|---|---|---|---|---|---|---|
| 3164.0109 | 0.0036 | −8822 | −0.0026 | 0.0019 | 1 | 4395.3354 | 0.0083 | −6871 | 0.0012 | −0.0012 | 1 |
| 3171.5847 | 0.0032 | −8810 | −0.0022 | 0.0024 | 1 | 4407.3262 | 0.0091 | −6852 | 0.0006 | −0.0021 | 1 |
| 3226.4891 | 0.0043 | −8723 | −0.0055 | −0.0001 | 1 | 6900.8873 | 0.0078 | −2901 | −0.0048 | 0.0000 | 2 |
| 3238.4789 | 0.0011 | −8704 | −0.0070 | −0.0014 | 1 | 6924.8697 | 0.0084 | −2863 | −0.0051 | −0.0004 | 2 |
| 3248.5774 | 0.0065 | −8688 | −0.0065 | −0.0008 | 1 | 6979.7828 | 0.0023 | −2776 | 0.0003 | 0.0048 | 2 |
| 3262.4621 | 0.0037 | −8666 | −0.0065 | −0.0006 | 1 | 6998.7105 | 0.0067 | −2746 | −0.0057 | −0.0014 | 2 |
| 3960.4868 | 0.0084 | −7560 | −0.0037 | 0.0026 | 1 | 7241.0657 | 0.0039 | −2362 | −0.0017 | 0.0000 | 2 |
| 3967.4282 | 0.0059 | −7549 | −0.0047 | 0.0015 | 1 | 7359.7189 | 0.0067 | −2174 | 0.0004 | 0.0003 | 2 |
| 3968.6886 | 0.0057 | −7547 | −0.0066 | −0.0004 | 1 | 7501.0946 | 0.0081 | −1950 | 0.0046 | 0.0020 | 2 |
| 3979.4182 | 0.0078 | −7530 | −0.0060 | 0.0001 | 1 | 7677.8088 | 0.0081 | −1670 | 0.0044 | −0.0012 | 2 |
| 3991.4075 | 0.0093 | −7511 | −0.0080 | −0.0021 | 1 | 8712.2167 | 0.0065 | −31 | 0.0011 | −0.0020 | 3 |
| 3998.3522 | 0.0085 | −7500 | −0.0058 | 0.0000 | 1 | 8714.1127 | 0.0041 | −28 | 0.0037 | 0.0007 | 3 |
| 4010.3427 | 0.0081 | −7481 | −0.0066 | −0.0010 | 1 | 8716.6337 | 0.0041 | −24 | 0.0001 | −0.0029 | 3 |
| 4022.3337 | 0.0095 | −7462 | −0.0069 | −0.0015 | 1 | 8718.5296 | 0.0047 | −21 | 0.0027 | −0.0002 | 3 |
| 4324.6507 | 0.0091 | −6983 | 0.0022 | 0.0015 | 1 | 8722.9465 | 0.0041 | −14 | 0.0018 | −0.0010 | 3 |
| 4328.4395 | 0.0080 | −6977 | 0.0042 | 0.0034 | 1 | 8727.3635 | 0.0023 | −7 | 0.0009 | −0.0018 | 3 |
| 4333.4873 | 0.0075 | −6969 | 0.0031 | 0.0022 | 1 | 8729.2595 | 0.0067 | −4 | 0.0035 | 0.0008 | 3 |
| 4352.4211 | 0.0037 | −6939 | 0.0032 | 0.0019 | 1 | 8731.7805 | 0.0025 | 0 | 0.0000 | −0.0026 | 3 |
| 4357.4697 | 0.0061 | −6931 | 0.0028 | 0.0013 | 1 | 8733.6764 | 0.0037 | 3 | 0.0026 | 0.0000 | 3 |
| 4362.5186 | 0.0068 | −6923 | 0.0028 | 0.0012 | 1 | | | | | | |

**Note.** 1-SuperWASP; 2-ASAS-SN and 3-TESS.

TESS and Gaia epoch photometry data using the Wilson–Devinney (WD) code (version 2015) to derive its parameters. The light curve analysis was initially performed using mode-2 (detached-configuration), however the solution converged in mode-3 (contact-configuration). The method adopted for modeling light curves is briefly discussed in Priya et al. (2013) and Joshi et al. (2016). The effective temperature of the primary ($T_h$) for the analysis was fixed at 6657 K, as adopted from estimates given in the Large sky Area Multi-Object Fiber Spectroscopic Telescope (LAMOST) survey.

Along with $T_h$, the parameters for hotter ($h$) and cooler ($c$) components, like the gravity darkening coefficients $g_h = g_c = 0.32$ (Lucy 1967), the bolometric albedos $A_h = A_c = 0.5$ (Rucinski 1969) and the limb darkening coefficient values $x_h = x_c = 0.63$ (Van Hamme & Wilson 2003), were fixed for the $I_{TESS}$ passband. The parameters such as the orbital inclination ($i$) of the system, the effective temperature of the secondary component ($T_c$), dimensionless potentials of the components ($\Omega_h = \Omega_c$) and relative luminosity of the primary component ($L_h$), were taken as adjustable parameters. Throughout the study, mass ratio is defined as $q = M_c/M_h$, where $M_c$ is mass of the cooler component and $M_h$ is mass of the hotter component. As no mass ratio was determined for V2840 Cygni in previous literature, a $q$-search was conducted to obtain the value of "$q$" (Rukmini et al. 2001; Priya et al. 2011, 2013, 2020). The adjustable parameters were iterated for a range of "$q$" values varying in the range from 0.1 to higher values in steps of 0.1, using the differential correction (DC) routine until a minimum residual was obtained at $q = 1.3$. The effect of third light ($l_3$) on the system was evaluated by taking it as an adjustable parameter. The DC routine was repeated and the best fit was obtained at $q = 1.357$. The best fit parameters derived from the analysis are as tabulated in Table 2 and the same values were used in the light correction (LC) routine to obtain the synthetic light curve of V2840 Cygni. The same procedure is followed using the Gaia photometric data and the best fit was obtained at $q = 1.325$. There is no effect due to third light in the Gaia photometric data analysis. The best fit light curve along with phase folded light curve from TESS and Gaia photometric observations is as shown in Figure 1.

### 3.2. Orbital Period Variation Study

To examine the dynamical changes in V2840 Cygni, the long term period variation study was performed for various epochs spanning over 15 yr of photometric data. For this investigation, the eclipse time variation ($O-C$) values were obtained using the derived ephemeris (Equation (1)) for V2840 Cygni. The derived ($O-C$) values are listed in Table 1 which are used to plot the $O-C$ diagram as displayed in Figure 2. A nonlinear least-squares method following a robust regression procedure, based on the Least Absolute Residuals (LAR) method, was applied from MATLAB's curve fitting toolbox package to obtain the best fit for ($O-C$) data. The observed ($O-C$) data are well within the 95% confidence bounds of the derived fit. The best fit ($O-C$) curve displays a parabolic trend superimposed by a cyclic variation indicating a secular

**Table 2**
Light Curve Solutions of V2840 Cygni using WD code

| Parameter | TESS Data without Third Light | TESS Data with Third Light | Gaia Data without Third Light |
|---|---|---|---|
| $q$ | 1.357 ± 0.001 | 1.357 ± 0.001 | 1.325 ± 0.012 |
| $T_c$ | 5653 ± 12 K | 5652±10 K | 5728 ± 9 K |
| $i$ | 57°.52 ± 0°.07 | 57°.67 ± 0°.12 | 57°.26 ± 0°.72 |
| $\Omega_h = \Omega_c$ | 4.3098 ± 0.0019 | 4.3049 ± 0.0036 | 4.3069 ± 0.0150 |
| $f(\%)$ | 11.53 ± 0.04 | 12.39 ± 0.04 | 12.04 ± 0.04 |
| $r_{h,pole}$ | 0.3309 ± 0.0001 | 0.3314 ± 0.0003 | 0.3281 ± 0.0017 |
| $r_{h,side}$ | 0.3465 ± 0.0002 | 0.3471 ± 0.0004 | 0.3428 ± 0.0020 |
| $r_{c,pole}$ | 0.3817 ± 0.0002 | 0.3817 ± 0.0004 | 0.3746 ± 0.0021 |
| $r_{c,side}$ | 0.4026 ± 0.0003 | 0.4026 ± 0.0005 | 0.3941 ± 0.0026 |
| $l_h$ | 6.8859 ± 0.0197 | 6.7559 ± 0.082 | 6.4208 ± 0.1400 |
| $l_c$ | 5.6805 | 5.7904 | 6.1456 |
| $l_3$ | ... | 0.0201 | ... |
| $\Sigma(O-C)^2$ | 0.0001 | 0.0001 | 0.0009 |

**Note.** $h$, $c$-hotter and cooler components.

decrease in the orbital period of V2840 Cygni. The best fit equation is given by

$$
\begin{aligned}
(O-C) = &-4.7350(\pm 0.4734) \times 10^{-10} \\
&\times E^2 - 4.4940(\pm 0.4818) \times 10^{-6} \times E \\
&-0.0037(\pm 0.0020) + 0.0105(\pm 0.0035) \\
&\times \sin(1.3111(\pm 0.0319) \times 10^{-3} \\
&+2.4940(\pm 0.4170).
\end{aligned} \tag{2}
$$

From the above equation, the rate of period decrease is determined as $\dot{P} = 5.48(\pm 0.58) \times 10^{-7}$ day yr$^{-1}$. The period decrease could be due to rapid mass transfer from the more massive to the less massive component of V2840 Cygni. Further, the cyclic variation is found to exhibit a semi-amplitude of ~0.0105 day and period of about 8.28 yr. A cyclic periodic variation is generally attributed to the presence of a third body causing the light travel time effect (LTTE) (Borkovits & Hegedüs 1996) or to the magnetic activity cycle, i.e., Applegate mechanism (Applegate 1992). The possible reason for the cyclic variability in $(O-C)$ for V2840 Cygni is discussed in detail in Section 6. The quadratic+sine fit is represented by a thick line along with confidence bounds appearing as dotted lines in the upper panel of Figure 2, where the residuals between the $(O-C)$ and the quadratic+sine fit plotted in the lower panel of the same figure show the goodness of fit. However, the results are currently suggestive but not conclusive and additional observations may help in confirming the same.

### 3.3. Spectral Line Analysis

The chromospheric activity on the binary components can be studied using the spectral line analysis (Barden 1985; Devarapalli & Jagirdar 2016). The normalized spectra of

V2840 Cygni with respect to the standard star is as depicted in Figure 3. The absorption line profiles suggest strong magnetic activity in the photosphere (such as stellar spots) or in the lower chromosphere as evident by filled-in emission or only emission respectively. In particular, earlier studies found that the filled-in effect in the H$\alpha$ line can be generally attributed to the ongoing chromospheric activity, which affects the depth of the H$\alpha$ absorption line which is extensively studied in the literature for over-contact binaries (Rukmini & Priya 2016; Devarapalli et al. 2020; Xia et al. 2021). The H$\alpha$ emission is considerably stronger in those components with deeper convective zones. Most of the components in CBs have similar temperatures with $\Delta T \leqslant 1000$ K and the orbital period equal to the rotational period but may differ in the depths of the convective zones, resulting in variable dynamo action (Vilhu & Walter 1987). The weaker absorption profiles of Balmer and Ca II H & K in addition to those in Mg, Na and G band are indicative of strong magnetic activity like prominences and flares which are extensions of magnetic activity from the photosphere into deep chromospheric layers. Whereas, the He I triplet reveals upper chromospheric activities such as strong flares, prominences, etc. (Drechsel et al. 1982; Montes & Martin 1998) which are further substantiated by ultraviolet (UV) emission and low energy X-ray observations. They also become very crucial in understanding the spatial distribution of magnetic activity when studied as a function of orbital phase (Devarapalli et al. 2020). The orbital phases for the observed spectra of V2840 Cygni (Section 2.3), were computed using the ephemeris (Equation (1)) and found to be lying between 0.53 and 0.73, the egress after the secondary eclipse. The spectra were collected in the spectral range 3800–8000 Å, to study the absorption profiles of Ca II K (3933.7 Å), Ca II H (3968.5 Å), H$\delta$ (4102.9 Å), H$\gamma$ (4341.7 Å), H$\beta$ (4862.7 Å), Na D (5895.6 Å) and H$\alpha$ (6564.6 Å), as shown in Figure 4. The variation in the observed spectral lines (Ca II H & K, Balmer, Na and Mg) in terms of their equivalent widths (EqWs) at the derived phases are being reported (Table 4) to check for the correlation with any ongoing stellar activity in V2840 Cygni. The deduced EqWs of the spectral lines at the different phases do not show any significant phase dependent variability, contrary to similar spectral-type binaries (Devarapalli et al. 2020), which can be due to screening of the variability by the contribution from both components throughout the orbital cycle. However, a likely chance of a close-in third body contaminating the binary spectra (Hendry & Mochnacki 1998) cannot be neglected.

## 4. Catalog of HMRCBs

HMRCBs with $q \geqslant 0.7$ stand-out as the testing tools for understanding the origin and evolution of CBs in light of the TRO model and are known to behave in high contrast to those of other CBs showing much less efficient energy transfer rates (Webbink 1979; Mochnacki 1981; Csizmadia & Klagyivik 2004). We have complied a catalog of 59 well-studied short-period HMRCBs with $P < 1$ day and $q \geqslant 0.7$. The catalog with binary parameters is presented in Table 8.

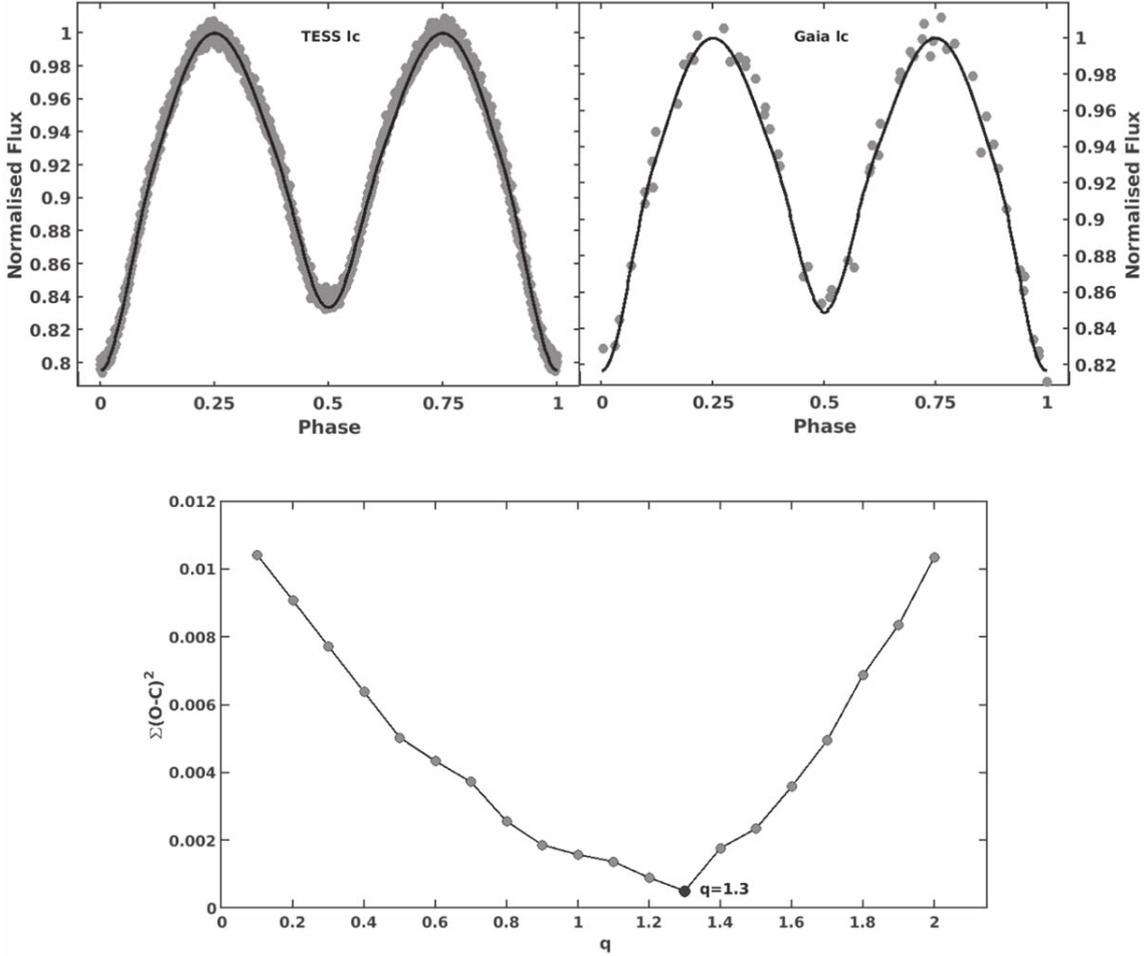

**Figure 1.** The best fit light curves (top) of V2840 Cygni for TESS and Gaia epoch photometric data and the $q$-search plot (bottom) featuring minimum $\Sigma(O - C)^2$ obtained at $q = 1.3$ as reported by the WD code.

The majority of the catalog (48 systems) has a degree of contact $\leqslant 25\%$ which is considered to be a weak-contact configuration. These cases fall in the range of spectral types A to M and 22% have derived third body parameters. Figure 5 shows the number density plots of $P$, $q$, $f\%$ and $T_{eff}$ with corresponding Gaussian fits for the cataloged HMRCBs. The distribution of periods has a minimum value of 0.1913 day and a maximum value of 0.8270 day with median at 0.2903 day. The parameters of the Gaussian fit are $\mu = 0.3290$ day, $\sigma = 0.1340$ day. The minimum and maximum values of the mass ratio are 0.70 and 1.40 respectively, with median at 0.90 where the critical value of HMRCBs is taken at $q \sim 0.70$ as per Csizmadia & Klagyivik (2004). The parameters of the bi-component Gaussian mixture fit are $\mu = 0.99$ and $\sigma = 0.22$. The fill-out factor values ($f\%$) of the binary systems are as shown in Figure 5 in which the minimum and maximum values are 0.01% and 90.30% respectively, with median at 12.70%. The distribution of effective temperatures has a minimum value 3460 K and a maximum value 8000 K with median of 5350 K. The parameters of the corresponding Gaussian fit are $\mu = 5287$ K and $\sigma = 965$ K. V2840 Cygni is an HMR, weak CB at extreme values of $P$ and $T_{eff}$ in comparison with other cataloged HMRCBs. The distribution can help in conclusive characterization of HMRCBs by increasing the sample size of the catalog.

The estimation of absolute parameters of CBs is significant in understanding the characteristic nature of CBs and evolution of their components. The masses ($\log M$) and radii ($\log R$) values of each component of the CB are estimated using the empirical relations given by Gazeas & Niarchos (2006) and

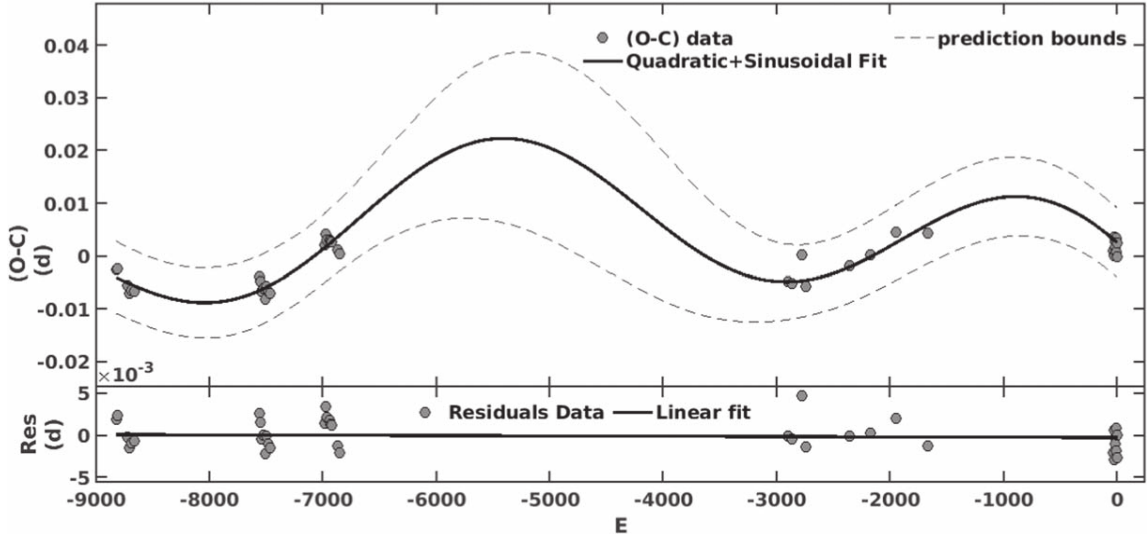

**Figure 2.** Top panel: $(O - C)$ data with best fit-quadratic and quadratic+sine functions; bottom panel: residuals for the $(O - C)$ fit.

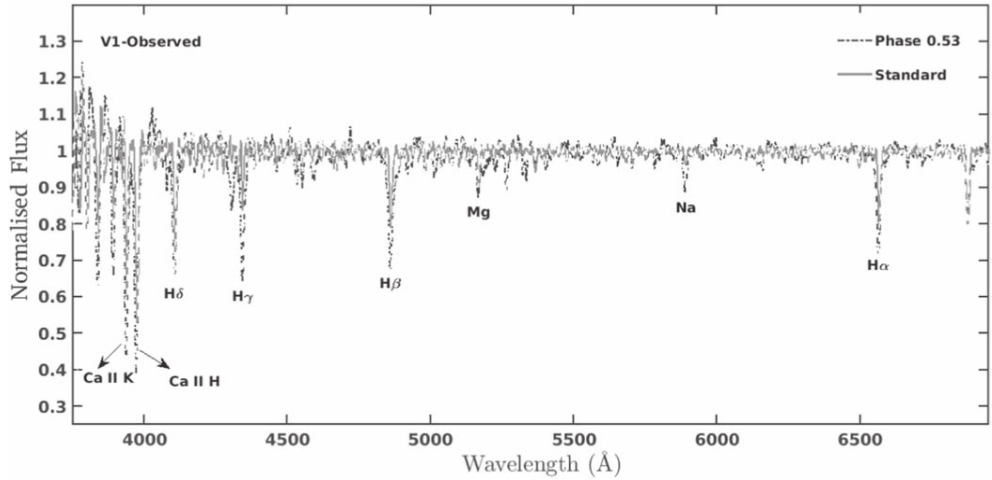

**Figure 3.** Comparison between normalized spectra of variable V2840 Cygni at phase 0.53 and standard HD 203454 with various line profiles, taken from HCT.

Djurašević et al. (2016) and used for further studies. The various findings with respect to the above are discussed in Section 6.

## 5. LAMOST Spectra of HMRCBs

LAMOST is a 4 m aperture telescope with a wide field of view (FOV) of $5° \times 5°$ (Cui et al. 2012). Its distinctive characteristic of taking 4000 spectra in a single exposure covering wavelength range 3690–9000 Å at a resolving power $R \sim 1800$ makes it an extraordinary and powerful instrument for spectroscopic studies. We collected 29 low-resolution spectra for 17 of our cataloged HMRCBs from the LAMOST

Data Release 7 (DR7) database portal.[3] V2840 Cygni was observed in 2013 by the LAMOST NEWCAM v2.0 with the observation median UTC 11:22:00, 2013 October 26 and its spectrum is displayed in Figure 6. The parameters of HMRCBs, including V2840 Cygni, provided in the LAMOST database are listed in Table 7. The epochs from literature were used to compute the respective phases of the LAMOST spectra. The identification and interpretation through visual inspection of the spectral line features, indicating the magnetic activity, are being reported for the first time in the current study.

The brief descriptions of all the spectra are discussed below:

---

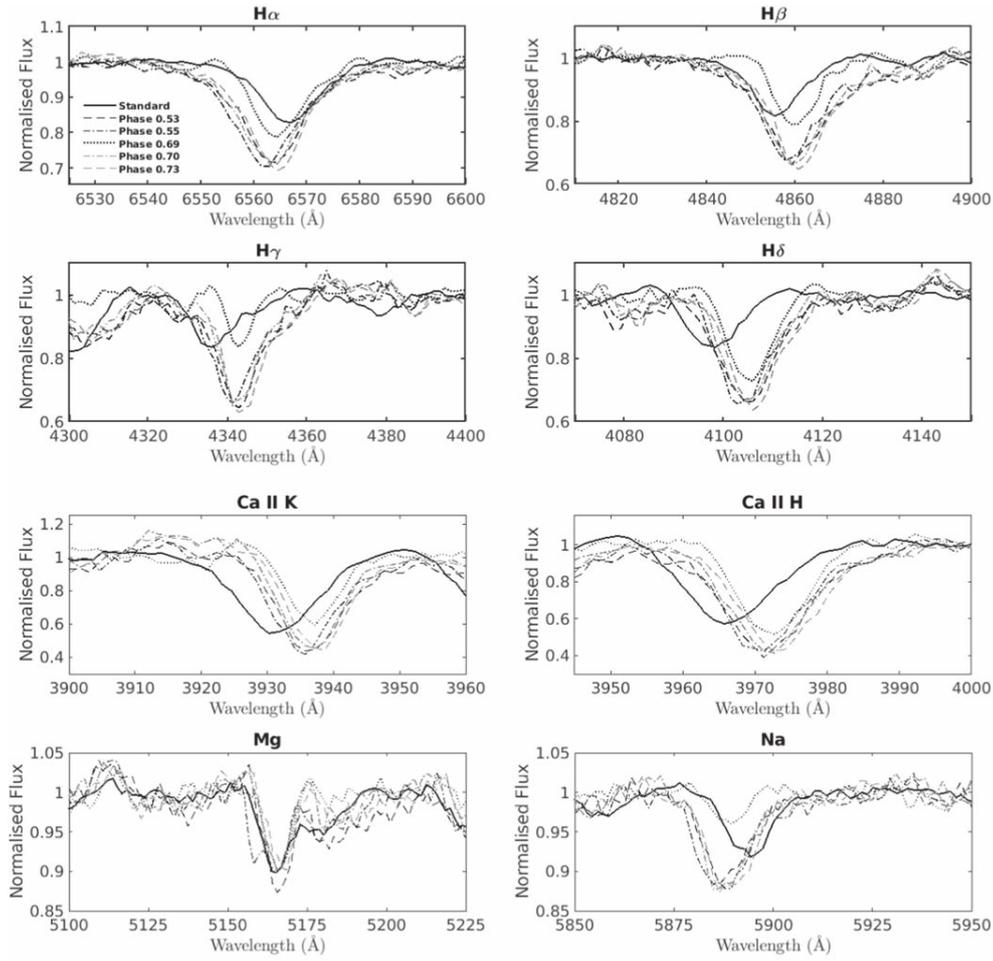

**Figure 4.** Balmer and metal line profiles of V2840 Cygni obtained at various orbital phases along with those of the standard HD 203454. The notation in all plots is the same as described in the first plot.

All the HMRCBs show Ca II IRT lines and the G band. Among these, ASAS J174406+2446.8, DZ Lyn, IK Boo, V1370 Tau and V1799 Ori exhibit relatively weaker Balmer lines. Whereas, BG Vul, DZ Lyn, FZ Ori, IK Boo, KIC 7950962, V1101 Her, V1370 Tau and V2840 Cygni display relatively weaker metal lines (Na doublet, Mg I b triplet, Ca II H & K and Ca II IRT lines). AU Ser, V1799 Ori and V523 Aur show relatively stronger absorption line profiles of only Na doublet and Mg I b triplet lines. Additional spectral features of individual variables are stated in the Notes column of Table 7. All the other spectral characteristics of HMRCBs are discussed in Section 6.

## 6. Discussion and Conclusions

We present the first detailed study of a neglected short-period eclipsing binary V2840 Cygni. The light curve solutions for the TESS and Gaia data were derived using the WD code. Due to the unavailability of mass ratio ($q$-values) through radial velocity observations, a $q$-search method was applied in the photometric analysis, and the system is found to belong to the W-subtype CB with an HMR ($q \sim 1.3$) and weak thermal contact ($\Delta T \sim 1000$ K). It shows a weak-contact configuration with a low fill-out factor ($f \sim 12\%$). The low amplitude of light variation in the light curve is attributed to the low inclination ($i \sim 57°$) of V2840 Cygni, indicating the partially eclipsing nature of the binary, which is similar to CSS J071813.2 +505000, NSVS 2459652, NSVS 7377875 (Kjurkchieva et al. 2017); KIC 9532219 (Lee et al. 2016); AA UMa (Lee et al. 2011); XY Leo, EE Cet and AQ Psc (Djurašević et al. 2006). Several studies show that late W-subtype systems are magnetically very active (Coughlin et al. 2008; Qian et al. 2014; Li et al. 2016; Mitnyan et al. 2018; Zhou & Soonthornthum 2019). However, V2840 Cygni displays no evidence of spots or flaring activity in the light curve. The absolute parameters of V2840 Cygni were determined using Gazeas ([2009](#)) as $M_h = 1.273 \pm 0.002\ M_\odot$, $M_c = 1.696 \pm 0.003\ M_\odot$, $R_h = 1.631 \pm 0.002\ R_\odot$, $R_c = 1.844 \pm 0.003\ R_\odot$, $L_h = 3.566 \pm 0.005\ L_\odot$ and $L_c = 4.569 \pm 0.007\ L_\odot$, suggesting the secondary component is slightly more massive than the primary. To understand its evolutionary state and characterize such HMRCBs, a catalog of well-studied HMRCBs (Section [4](#)) has been compiled (Table [8](#)). The mass–radius relation is plotted in Figure [7](#) for the cataloged HMRCBs including V2840 Cygni. The zero age main sequence (ZAMS) and terminal age main sequence (TAMS) lines are extracted from Stepien ([2006](#)). It is observed that the primary component of V2840 Cygni lies on the TAMS line and the secondary component falls between ZAMS and TAMS, suggesting the primary to be more evolved or moderately evolved than the secondary for their main sequence masses. Also, a significant fraction of HMRCB secondaries are found to be overluminous for their main sequence masses, compared to their corresponding primary masses as seen from the log $M$ versus log $R$ relation (Figure 7).

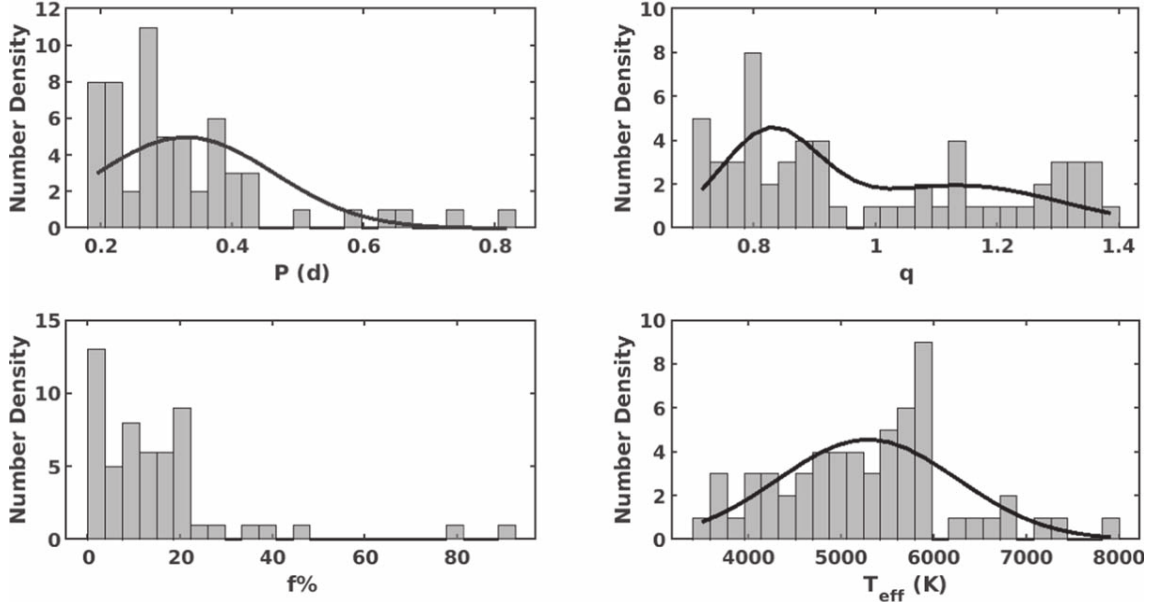

**Figure 5.** Number density plots of cataloged HMRCBs for period, mass ratio, fill-out factor and effective temperature. The solid line represents the corresponding Gaussian fit.

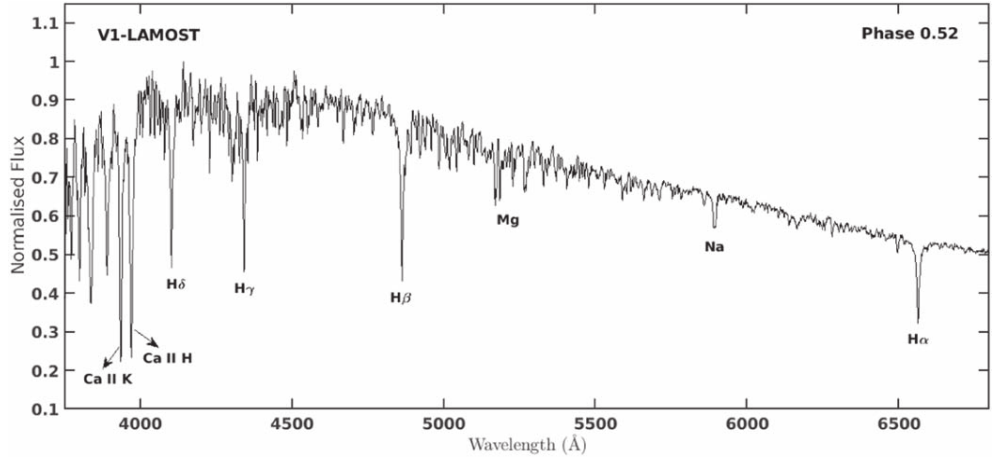

**Figure 6.** LAMOST spectrum of V2840 Cygni observed at Phase 0.52.

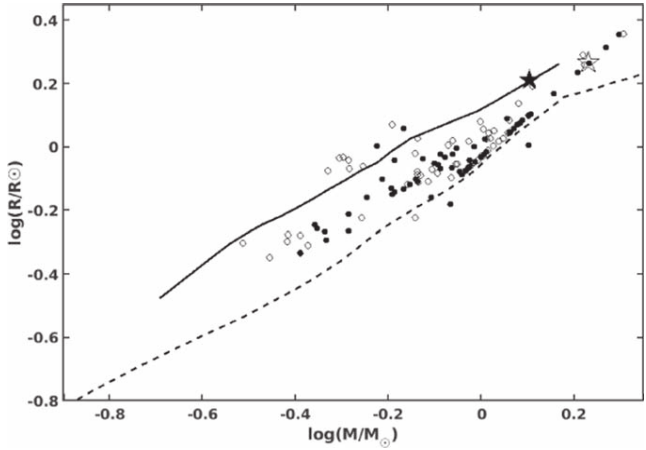

**Figure 7.** log $M$–log $R$ diagram-the primary (filled pentagram) and the secondary (open pentagram) of V2840 Cygni along with cataloged HMRCB primary components (filled circles) and secondary components (open circles). The ZAMS (dashed line) and TAMS (solid line) are adopted from Stepien (2006).

From period variation study on data spanning over 15 yr, it is found that the system's period is observed to be decreasing at a rate of $\sim 10^{-7}$ day yr$^{-1}$, which could be due to a rapid conservative mass transfer from the more massive secondary to the less massive primary component. Considering a conservative mass transfer in V2840 Cygni, the mass transfer rate is measured from the observed period variation and is about $\dot{M} = 1.48 \times 10^{-6}\ M_\odot$ yr$^{-1}$. We have determined two timescales for the primary component, one to fill its Roche Lobe ($t_{RL}$) and the other to gain mass ($t_{MT}$) from the secondary component for the observed mass transfer rate, using the following relations (Wang 1999)

$$t_{RL} = f(q) \times t_{MT},\qquad(3)$$

and

$$t_{MT} = \frac{M_c}{\dot{M}_c},\qquad(4)$$

where $\dot{M}_c$ is the rate of mass transfer and $f(q)$ is equal to the ratio of two timescales and is a function of the mass ratio which is measured using the following relation

$$f(q) = \left[ \frac{2(1-q)}{q} - \frac{2r_L}{3q^{1/3}} \right. $$
$$\left. \times \left[ \frac{1}{1+q^{1/3}} - 2\ln(1+q^{-1/3}) \right](1+q) \right]^{-1}].\quad(5)$$

Following the above equations, the value obtained for $f(q)$ is 1.15(>0), which shows the shrinking of the secondary component's Roche lobe with decreasing $q$ and also indicates the stability of the mass transfer between the components since $|f(q)| > 1$ as defined by Wang (1999). Though the mass transfer is stable, when compared, the mass transfer timescale $t_{MT}$ ($\sim 10^6$ yr) is relatively shorter than the thermal timescale $t_{Th}$ ($\sim 10^7$ yr) (Paczynski 1971), suggesting that the mass transferring secondary component cannot maintain its thermal equilibrium with the primary (as evident by $\Delta T$ in the observed light curve solution).

In addition to the period decrease observed in the $(O - C)$ diagram, the superimposed cyclic variation on the long-term quadratic period change may be attributed to magnetic activity cycle or presence of a third body, or both. The magnetic activity cycles are commonly observed in active binary systems such as TY Boo (Yang et al. 2007); and VW Cep (Mitnyan et al. 2018). To verify the same in V2840 Cygni, the quadruple moment of both the components is determined using the relation given by Lanza & Rodonò (2002)

$$\Delta Q = -\frac{\Delta P}{P} \times \frac{Ma^2}{9},\qquad(6)$$

where $\Delta P$ is observed orbital period variation, $M$ is the mass of the active star and $a$ is the semimajor axis of the binary orbit. The quadruple moments for both the components of V2840 Cygni are calculated to be $\Delta Q_h = 3.8 \times 10^{45}$ g cm$^2$ and $\Delta Q_c = 5.1 \times 10^{45}$ g cm$^2$, which are much less than those of typical values for active stars ($10^{51-53}$ g cm$^2$, Lanza & Rodono 1999). Thus, the quadruple moment of either of the components of V2840 Cygni is insufficient to produce and cause magnetic activity cycles, indicating that the Applegate mechanism may not be a feasible explanation for the observed cyclic period variation. However, the cyclic variation in the orbital period can be reasonably explained via the presence of a third body. Many such binaries are known to harbor third bodies (Fang et al. 2019; Devarapalli et al. 2020; Zhu et al. 2021), which can be detected by their contribution to total light in the light curve analysis using the WD method (Zhang et al. 2020), and through cyclic variation observed in period variation studies. The same was verified with the significant contribution of third light ($\sim 0.102$) in the light curve analysis of V2840 Cygni. The parameters of an additional component were also derived from period variation study, using the following relation for mass function (Borkovits & Hegedüs 1996)

$$f(m_3) = \frac{m_3^3 \sin^3}{(M_{bin} + m_3)} = \frac{4\pi^2 a_{12}^3 \sin^3 i}{GP_3^2},\qquad(7)$$

where $M_{bin}$ and $m_3$ are the total mass of the binary and the third body respectively, and $G$ is the gravitational constant. The derived third body parameters are expressed in Table 3. The third body possibly is a low-mass red dwarf with luminosity $\sim 0.067\ L_\odot$, signifying it to be fainter than the host binary, orbiting with a period $\sim 8$ yr at a distance of $\sim 5.28$ au from the barycenter of the system. The close proximity of the third body may play a key role in the evolution of the CB (Zakirov [2010]). Since AML and the Applegate mechanism are incompatible for V2840 Cygni, the plausible explanation for the period variation is the existence of a third body along with conservative mass transfer, via TRO.

**Table 3**
Parameters Derived from the Orbital Period Variation Study

| Parameter | Value | Parameter | Value | Parameter | Value |
|---|---|---|---|---|---|
| $\dot{P}$ | $-5.48(\pm 0.59) \times 10^{-7}$ day yr$^{-1}$ | $P_3$ | $8.28(\pm 0.20)$ yr | $a_{12} \sin i$ | $0.91(\pm 0.30)$ au |
| $\dot{M}$ | $1.48(\pm 0.20/) \times 10^{-6}$ $M_\odot$ yr$^{-1}$ | $f(m_3)$ | $0.0109(\pm 0.0005)$ $M_\odot$ | $a_3$ | $5.28(\pm 3.60)$ au |
| $\dot{H}_{\text{orb}}(\times 10^{42})$ | $2.63 \pm 0.04$ cgs | $m_{3(\text{min})}$ | $0.51(\pm 0.18)$ $M_\odot$ | | |

**Table 4**
EqWs of Prominent Spectral Lines in the Observed Spectra of V2840 Cygni

| Phase | H$_\alpha$ (6563 Å) | H$_\beta$ (4861 Å) | H$_\gamma$ (4340 Å) | H$_\delta$ (4102 Å) | Na (5890 Å) | Ca II K (3934 Å) | Ca II H (3968 Å) | Mg (5169 Å) |
|---|---|---|---|---|---|---|---|---|
| 0.53 | $3.39 \pm 0.21$ | $4.20 \pm 0.24$ | $2.98 \pm 0.16$ | $4.90 \pm 0.24$ | $1.50 \pm 0.28$ | $6.96 \pm 0.20$ | $7.55 \pm 0.22$ | $1.20 \pm 0.16$ |
| 0.55 | $3.86 \pm 0.23$ | $3.88 \pm 0.22$ | $3.32 \pm 0.19$ | $4.30 \pm 0.19$ | $1.54 \pm 0.21$ | $6.25 \pm 0.17$ | $6.82 \pm 0.20$ | $1.15 \pm 0.17$ |
| 0.69 | $3.96 \pm 0.23$ | $4.14 \pm 0.21$ | $2.66 \pm 0.17$ | $4.39 \pm 0.19$ | $1.02 \pm 0.12$ | $6.81 \pm 0.20$ | $7.39 \pm 0.22$ | $0.83 \pm 0.16$ |
| 0.70 | $3.51 \pm 0.21$ | $4.43 \pm 0.21$ | $3.54 \pm 0.15$ | $3.44 \pm 0.17$ | $1.55 \pm 0.21$ | $6.60 \pm 0.18$ | $7.22 \pm 0.20$ | $1.10 \pm 0.15$ |
| 0.73 | $3.28 \pm 0.19$ | $3.58 \pm 0.19$ | $3.30 \pm 0.15$ | $4.40 \pm 0.20$ | $1.58 \pm 0.21$ | $6.08 \pm 0.17$ | $7.20 \pm 0.20$ | $1.35 \pm 0.17$ |

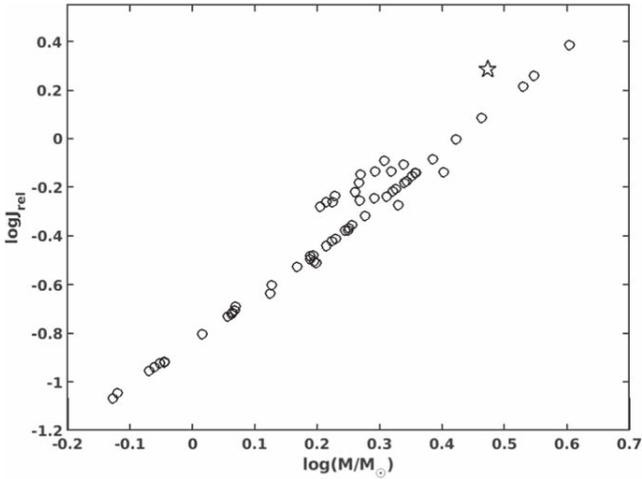

**Figure 8.** log $M$ vs. log $J_{\text{rel}}$ of HMRCBs. V2840 Cygni is represented by an open pentagram.

More than forty percent of the HMRCBs of our catalog have been studied for period variation in literature. A fraction of them showed period decrease which was attributed to the mass/energy transfer plus AML from the system either due to stellar winds or strong magnetic activity, or the influence of a third body or the combined effects. The other fraction showed an increasing period variation suggesting their evolution into a broken contact phase through TRO. HMRCBs with larger orbital angular momentum are suggestive of evolution through TRO, whereas those with very small orbital angular momentum signify the existence of past episodes of AML during their evolution. From period variation rates ($\dot{P}$), we can understand the timescales at which both TRO and AML progress.

When the relative orbital angular momentum ($J_{\text{rel}}$) is studied for mass of the binary system ($M_{\text{bin}}$), it becomes a good indicator in relating stability with evolution of the system. The $J_{\text{rel}}$ for V2840 Cygni along with other HMRCBs is derived using the following relation (Popper & Ulrich [1977])

$$J_{\text{rel}} = M_h M_c \left[ \frac{P}{(M_h + M_c)} \right]^{1/3}, \qquad (8)$$

where period $P$ is in days and component masses $M_{c,h}$ in solar units. The derived log $J_{\text{rel}}$ values of most of the HMRCBs (see Table 8) are correlated with those defined for the CBs by Popper & Ulrich ([1977]), with values greater than but closer to $-0.5$. The systems with greater log $J_{\text{rel}}$ and period increase tend to evolve into broken-contact phase and those with period decrease tend to evolve into much a deeper contact phase, i.e., at the upper right end of Figure 8. Those with lower log $J_{\text{rel}}$ values with period decrease tend to evolve into over-contact or deep contact phase, i.e., at the lower left end of Figure 8 and vice-versa. The systems which appear near the extreme lower left end might have undergone AML earlier and could be progenitors to mergers. The same can be appropriately understood when correlated with period variations, in light of TRO-controlled or AML-controlled evolution.

The graph of log $M$ versus log $J_{\text{rel}}$ is plotted for all HMRCBs including V2840 Cygni (in Figure 8), and the variable V2840 Cygni is observed to exhibit a relatively higher value of log $J_{\text{rel}}$,

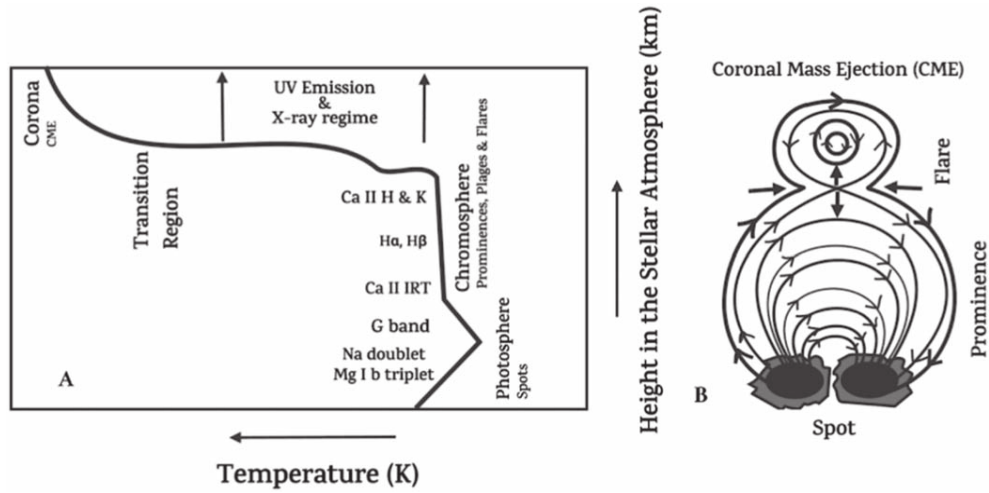

**Figure 9.** (A) The height vs. temperature profile representing various atmospheric layers with features and corresponding spectral lines formed in those regions of the stellar atmosphere depicted for an active star as a function of height and temperature. The height and temperature axis values are not included due to variability over a range of spectral classes of the cataloged binary's components. (B) A schematic diagram of magnetic activity, relating features such as spot, prominence, flare and CME.

**Table 5**
Plausible HMRCB Evolutionary Scenarios Based on Current Period Variation and Contact Configuration

| Model | | Criterion and Scenario as in example HMRCB |
|---|---|---|
| | Case 1 | Period Increase+weak contact configuration-Marginal contact to Broken contact configuration as in V2840 Cygni (Present Study) and V366 Cas (Yang et al. 2013) |
| | Case 2 | Period Decrease+weak contact configuration-Progenitor of much deeper Contact configuration: Semidetached to Marginal contact configuration as in ASAS J174406+2446.8 (Shi et al. 2020); Marginal to Deep contact configuration as in IK Boo (Kriwatta-nawong et al. 2017a). |
| TRO ($\log J_{rel} \geqslant -0.5$) | Case 3[a] | Period Increase+over contact or deep contact configuration-Over contact/Deep contact to Marginal contact configuration (a failed merger oscillating around a critical mass ratio due to variable AML while never breaking Contact configuration) as in UZ CMi (Qian et al. 2013b) |
| | Case 4[a] | Period Decrease+weak/over contact-Marginal/Over contact to Over contact/Deep contact configuration (a failed merger oscillating around a critical mass ratio due to variable AML) as in V781 Tau (Li et al. 2016); V2284 Cyg (Wang et al. 2017) and GDS J1056047-604149 (Li et al. 2022) |
| | Case 1 | Period Increase+weak contact configuration-Progenitor of Contact configuration via Magnetic braking as in AD Cnc (Qian et al. 2007) |
| | Case 2 | Period Decrease+weak contact configuration-Current state to deeper Contact configuration through magnetic braking/AML via stellar winds or LTTE via presence of third body as in BI Vul (Qian et al. 2013a) and EH CVn (Xia et al. 2018) |
| AML ($\log J_{rel} < -0.5$) | Case 3[a] | Period Increase+Over contact/Deep contact configuration -a failed merger oscillating around critical mass ratio, may be due to variable AML as in FI Boo (Christopoulou & Papageorgiou 2013) |
| | Case 4 | Period Decrease+Over contact/Deep contact configuration-Progenitor of merger as in 1SWASP J133105.91+121538.0 (Lohr et al. 2012) |

**Note.**
[a] These scenarios are distinct suggestive possibilities and open for further validation.

affirming its conservative nature and its evolved or moderately evolved binary phase, further evolving into a much deeper CB state as predicted by the TRO model. Such a rarely observed configuration is further substantiated through long-term high-precision photometric and spectroscopic observations.

Using the above results, a possible correlation is compiled between observed period variation and contact configuration for relative $\log J_{rel}$ (Popper & Ulrich 1977) values. The attempt was to shed light on the possible key stages of evolution such as exchange of mass between the components or AML from the system. We have considered TRO-controlled and AML-controlled evolution models around the limit $\log J_{rel} \sim -0.5$, as listed in Table 5. The distinct cases could represent a combination of different phases of binary evolution.

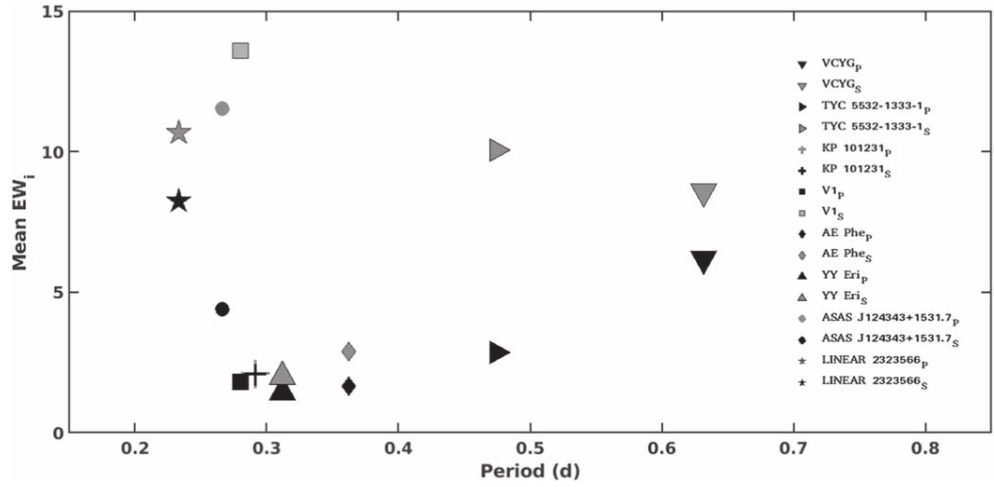

**Figure 10.** Period vs. mean intrinsic equivalent widths (EW$_i$) of the Hα line for CBs (marginal, over-contact and deep)- V2840 Cygni (Present Study); TYC 5532-1333-1 (Devarapalli et al. 2020); KP 101231 (Devarapalli & Jagirdar 2016); V2840 Cygni (Rukmini & Priya 2016); AE Phe (Vilhu & Maceroni 2005); YY Eri (Vilhu & Maceroni 2005); ASAS J124343+1531.7 (Xia et al. 2021) and LINEAR 2323566 (Xia et al. 2021).

**Table 6**
Significant Characteristic Features of HMRCBs from Literature

| Variable | Key Results/Remarks |
| --- | --- |
| VZ Psc | It showed strong Ca II emission associated with each of the two components of the system (Hrivnak & Milone 1985). |
| V574 Lyr | The spectroscopic study showed weak emission lines of Hα, Hβ, Ca II H & K and IRT in the subtracted spectra of the variable from synthesized spectra, inferring it to be an active EW variable. It also displayed starspots (Long et al. 2019). |
| SW Lac | The PSPC spectrum indicates significant low-energy absorption, $n_H \sim (15–65) \times 10^{19}$ cm$^{-2}$. But for the cool component, emission is poorly constrained. Comparing ROSAT and Einstein observations, the ratio of mean X-ray luminosities is 60%–80% brighter in ROSAT. The high $(fXP)_{cool}$ value for SW Lac is a consequence of it being the most X-ray luminous object in the sample (McGale et al. 1996). |
| OO Aql | OO Aql systems probably resulted from mergers of slightly evolved, detached binary systems. This type of system can be considered a subclass of CBs characterized by unusually low effective temperature for the observed large dimensions (Lu & Rucinski 1993). |
| KIC 7950962 | Key atmospheric parameters are listed in the catalog Zhang et al. (2019). |

The observed characteristics of V2840 Cygni are similar to those of ASAS J174406+2446.8 (Shi et al. 2020), the HMRCB that has also been studied with reference to the TRO model. Thus, results from all the studies on V2840 Cygni strengthen the TRO model while signifying the HMRCB stage to be a remarkable episode of the same.

In spectral studies, the variability observed in the strength of the spectral lines can be attributed to various magnetic activities such as spots, prominences, flares, plages, coronal mass ejections (CMEs), etc. occurring in various layers of the active stellar atmosphere, as schematically illustrated in Figure 9. The spectroscopic study of V2840 Cygni was conducted to understand the photometrically elusive magnetic activity in the system. The spectral study done on our observations and

LAMOST indicate Balmer lines along with the metal lines (Figure 4) show filled-in absorption profiles at various phases, however, with weakly phase-dependent variation. The spectral types indicate that both the components have an outer-convective envelope with an active chromosphere. The asymmetry in the absorption profiles could either be due to chromospheric flares, winds, etc. (Vilhu et al. 1991) or due to photospheric absorption (Cram & Mullan 1985; Rukmini & Priya 2016). To validate the surface activity of both the components and signature of close-in third body in the spectra, continuous high resolution spectroscopic observations, covering all the orbital phases, are strongly needed (Hendry & Mochnacki 1998).

From Figure 10, we can observe that the difference in mean intrinsic equivalent widths (EW$_i$) of Hα for primary and secondary is relatively high for over-contact and deep CBs which could be due to highly variable dynamo action. This is in contrast with the weak-contact systems, which exhibited

**Table 7**
Parameters of HMRCBs Provided in the LAMOST Database

| HMRCB | Date | Survey | $T_{\rm eff}$ (K) | RV (km s$^{-1}$) | Fe/H | log $g$ (cgs) | Notes |
|---|---|---|---|---|---|---|---|
| 1SWASP J133105.91+121538.0 | 2016/01/18 | IV | 4852 | −9.86 | −0.163 | 4.724 | |
| | 2016/02/06 | IV | 4872 | −5.62 | −0.138 | 4.729 | |
| | 2019/02/10 | VII | 4852 | −8.96 | −0.262 | 4.664 | |
| ASAS J174406+2446.8 | 2013/06/03 | I | 5493 | −21.19 | 0.195 | 4.021 | H$\alpha$ and Ca H & K lines show core filled-in, H$\beta$ and H$\delta$ are emission lines. |
| AU Ser | 2016/02/26 | IV | 5379 | −57.19 | 0.376 | 4.323 | All Balmer lines show core filled-in. |
| | 2016/03/29 | IV | 5309 | −86.84 | 0.335 | 4.367 | |
| BG Vul | 2014/11/02 | III | NA | NA | NA | NA | |
| | 2016/10/13 | V | 5985 | −0.32 | −0.197 | 4.056 | |
| DZ Lyn | 2016/12/17 | V | 6931 | −77.35 | −0.237 | 4.166 | |
| FP Lyn | 2014/11/20 | III | 5920 | −4.64 | 0.256 | 4.179 | |
| | 2014/12/08 | III | 5911 | −1.77 | 0.214 | 4.208 | |
| | 2014/12/26 | III | 5892 | −3.35 | 0.188 | 4.141 | |
| FV Cvn | 2013/04/20 | I | 5495 | −14.89 | 0.017 | 4.343 | H$\alpha$, H$\beta$ and Ca II H & K show core filled-in. |
| | 2014/01/17 | II | 5566 | −21.42 | 0.044 | 4.389 | |
| | 2014/03/25 | II | 5610 | −18.83 | 0.056 | 4.458 | |
| | 2015/04/07 | III | 5260 | −24.43 | −0.152 | 4.254 | |
| | 2017/02/16 | V | 5386 | −9.48 | −0.099 | 4.309 | |
| FZ Ori | 2017/03/14 | V | 6209 | 19.22 | −0.261 | 4.07 | H$\alpha$ and Ca II K show core filled-in. |
| IK Boo | 2016/05/18 | IV | 5628 | −54.52 | −0.6 | 4.125 | Ca II H shows core filled-in. |
| KIC 7950962 | 2015/09/25 | IV | 6841 | −59.19 | 0.216 | 4.173 | |
| PY Vir | 2014/03/10 | II | 5235 | −26.72 | −0.086 | 4.43 | |
| | 2017/05/13 | V | 5225 | −37.48 | −0.043 | 4.554 | |
| V1101 Her | 2014/04/20 | II | 6010 | −1 | −0.007 | 4.209 | |
| V1370 Tau | 2015/12/21 | IV | 5771 | −40.36 | −0.511 | 4.135 | |
| V1799 Ori | 2012/12/03 | I | 4942 | 7.15 | −0.126 | 4.011 | |
| V523 Aur | 2015/02/18 | III | 5297 | 20.28 | 0.104 | 4.42 | H$\alpha$ shows partly core filled-in. |
| V783 And | 2012/11/30 | I | NA | NA | NA | NA | All Balmer and Ca II H & K are emission lines |
| | 2014/12/19 | III | 4196 | −27.48 | −0.359 | 4.493 | |
| V2840 Cygni | 2013/10/26 | II | 6657 | −17.84 | 0.212 | 4.168 | |

**Note.** NA-Not Available.

**Table 8**
Catalog of HMRCBs Showing Parameters of all HMRCBs Collected from the Literature and the Current Work for V2840 Cygni

| Variable | $P$ (days) | $q$ | Method[*] | $f\%$ | $T_h$ (K) | $T_c$ (K) | $\log M_h$ ($M_\odot$) | $\log M_c$ ($M_\odot$) | $\log R_h$ ($R_\odot$) | $\log R_c$ ($R_\odot$) | $\log J_{rel}$ | $\dot{P}$ ($10^{-7}$ day yr$^{-1}$) | $M_3$ ($M_\odot$) | Reference |
|---|---|---|---|---|---|---|---|---|---|---|---|---|---|---|
| V2150 Cyg | 0.5919 | 0.80 | S | 21.00 | 8000 | 7920 | 0.207 | 0.110 | 0.236 | 0.194 | 0.087 | ⋯ | ⋯ | Mitnyan et al. (2020) |
| TYC 2675-663-1 | 0.4224 | 0.81 | S | ⋯ | 6480 | 5546 | 0.101 | 0.101 | 0.007 | 0.099 | −0.136 | ⋯ | ⋯ | Caballero-García et al. (2010) |
| OO Aql | 0.5068 | 0.84 | S | 21.40 | 5700 | 5472 | 0.157 | 0.081 | 0.170 | 0.138 | −0.001 | ⋯ | ⋯ | İçli et al. (2013) |
| FT UMa | 0.6547 | 0.98 | S | 17.40 | 7178 | 7003 | 0.232 | 0.223 | 0.264 | 0.260 | 0.218 | ⋯ | ⋯ | Yuan (2011) |
| SW Lac | 0.3207 | 1.28 | S | 1.34 | 5390 | 5170 | −0.092 | 0.015 | −0.056 | −0.010 | −0.179 | ⋯ | ⋯ | Şenavci et al. (2011) |
| V1799 Ori | 0.2903 | 1.34 | Tp | 3.50 | 4848 | 4781 | −0.140 | −0.015 | −0.102 | −0.048 | −0.233 | 0.18 | ⋯ | Liu et al. (2014) |
| V574 Lyr | 0.2731 | 0.95 | Pp | 10.00 | 4929 | 4690 | −0.042 | −0.064 | −0.087 | −0.097 | −0.377 | ⋯ | 0.03 | Long et al. (2019) |
| KIC 9532219 | 0.1982 | 1.20 | Pp | ⋯ | 5203 | 5031 | −0.066 | −0.143 | −0.180 | −0.222 | −0.509 | 5.30 | 0.09 | Lee et al. (2016) |
| DY CVn | 0.2459 | 1.25 | Pp | 13.20 | 4410 | 4297 | −0.224 | −0.131 | 0.004 | −0.090 | −0.600 | ⋯ | 0.11 | ZhiNing et al. (2017) |
| V1677 Sco | 0.2298 | 1.10 | Pp | 16.50 | 4317 | 4126 | −0.213 | −0.255 | −0.101 | −0.060 | −0.703 | −4.26 | 0.15 | Fang et al. (2019) |
| ASAS J174406+2446.8 | 0.3782 | 1.14 | Pp | 10.40 | 5492 | 5049 | 0.008 | 0.063 | 0.025 | 0.049 | −0.105 | 0.10 | 0.21 | Shi et al. (2020) |
| IK Boo | 0.3031 | 1.14 | Pp | 2.20 | 5781 | 5422 | −0.063 | −0.007 | −0.064 | −0.040 | −0.251 | −2.20 | 0.21 | Kriwattanawong et al. (2017b) |
| BI Vul | 0.2518 | 1.04 | Pp | 8.80 | 4600 | 4481 | −0.125 | −0.143 | −0.036 | −0.019 | −0.523 | −0.95 | 0.30 | Qian et al. (2013a) |
| V366 Cas | 0.7293 | 0.89 | Pp | 38.70 | 5860 | 5907 | 0.269 | 0.219 | 0.314 | 0.292 | 0.261 | 5.60 | 0.42 | Yang et al. (2013) |
| CSTAR 38663 | 0.2671 | 1.12 | Pp | 10.60 | 4616 | 4352 | −0.088 | −0.138 | −0.021 | 0.028 | −0.480 | ⋯ | 0.63 | Qian et al. (2014) |
| VZ Psc | 0.2613 | 0.80 | Pp | 90.30 | 4500 | 4118 | −0.186 | −0.284 | −0.041 | −0.067 | −0.688 | ⋯ | 0.67 | Ma et al. (2018) |
| AD Cnc | 0.2827 | 1.30 | Pp | 8.30 | 5000 | 4790 | −0.138 | −0.024 | −0.109 | −0.060 | −0.259 | 4.90 | 0.76 | Qian et al. (2007) |
| PY Vir | 0.3112 | 0.77 | Pp | 0.30 | 4830 | 4702 | 0.006 | −0.106 | −0.022 | −0.070 | −0.354 | ⋯ | 0.79 | Zhu et al. (2013a) |
| USPEB6 | 0.1913 | 0.70 | Pp | 0.30 | 3694 | 3714 | −0.358 | −0.513 | −0.245 | −0.302 | −1.067 | ⋯ | ⋯ | Priya et al. (2020) |
| AU Ser | 0.3865 | 0.71 | Pp | 19.80 | 5495 | 5153 | 0.077 | −0.072 | 0.071 | 0.007 | −0.236 | ⋯ | ⋯ | Gürol (2005) |
| CW Cmi | 0.3132 | 0.71 | Pp | 0.01 | 5610 | 4830 | 0.011 | −0.138 | −0.014 | −0.078 | −0.377 | 3.30 | ⋯ | Panpiboon et al. (2019) |
| NSVS 908513 | 0.3996 | 0.71 | Pp | 14.60 | 5923 | 5615 | 0.088 | −0.062 | 0.085 | 0.021 | −0.214 | ⋯ | ⋯ | Kjurkchieva et al. (2016) |
| V783 And | 0.2091 | 0.71 | Pp | 20.00 | 4200 | 3955 | −0.108 | −0.258 | −0.158 | −0.223 | −0.634 | ⋯ | ⋯ | Kjurkchieva (2021) |
| NSVS 2244206 | 0.2807 | 0.74 | Pp | 26.00 | 5157 | 4702 | −0.025 | −0.159 | −0.060 | −0.117 | −0.439 | ⋯ | ⋯ | Kjurkchieva et al. (2016) |
| USPEB1 | 0.2054 | 0.74 | Pp | 1.10 | 3460 | 3494 | −0.285 | −0.416 | −0.212 | −0.275 | −0.915 | ⋯ | ⋯ | Priya et al. (2020) |
| USPEB8 | 0.2052 | 0.74 | Pp | 0.30 | 4051 | 3984 | −0.286 | −0.417 | −0.263 | −0.298 | −0.917 | ⋯ | ⋯ | Priya et al. (2020) |
| NSVS 2700153 | 0.2285 | 0.78 | Pp | 7.10 | 4785 | 4689 | −0.187 | −0.298 | −0.142 | −0.032 | −0.719 | ⋯ | ⋯ | Dimitrov & Kjurkchieva (2015) |
| VSX J062624.4+570907 | 0.2806 | 0.78 | Pp | 16.20 | 5350 | 5044 | −0.027 | −0.136 | −0.064 | −0.111 | −0.421 | ⋯ | ⋯ | Kjurkchieva et al. (2016) |
| GQ Cancri | 0.4222 | 0.80 | Pp | ⋯ | 5250 | 5247 | 0.101 | 0.004 | 0.099 | 0.058 | −0.139 | ⋯ | ⋯ | Samec et al. (2017) |
| V1101 Her | 0.3827 | 0.80 | Pp | 14.20 | 5920 | 5690 | 0.070 | −0.027 | 0.060 | 0.018 | −0.204 | ⋯ | ⋯ | Pi et al. (2017) |
| 1SWASP J133105.91 +121538.0 | 0.2180 | 0.81 | Pp | ⋯ | 4977 | 4677 | −0.078 | −0.138 | −0.032 | −0.087 | −0.501 | −40.00 | ⋯ | Lohr et al. (2012) |
| NSVS 2729229 | 0.2288 | 0.81 | Pp | 25.00 | 3942 | 3692 | −0.194 | −0.285 | −0.130 | −0.040 | −0.714 | ⋯ | ⋯ | Kjurkchieva et al. (2018) |
| NSVS 8626028 | 0.2174 | 0.81 | Pp | 20.70 | 4318 | 4095 | −0.246 | −0.330 | −0.158 | −0.074 | −0.802 | ⋯ | ⋯ | Dimitrov & Kjurkchieva (2015) |
| QQ Boo | 0.2765 | 0.83 | Pp | 10.30 | 5789 | 5392 | −0.034 | −0.115 | −0.074 | −0.108 | −0.411 | ⋯ | ⋯ | Rahimi et al. (2021) |
| V1370 Tau | 0.2955 | 0.83 | Pp | 11.00 | 5634 | 5949 | −0.013 | −0.094 | −0.047 | −0.082 | −0.366 | ⋯ | ⋯ | Rahimi et al. (2021) |
| FZ Ori | 0.4000 | 0.86 | Pp | 2.00 | 5940 | 5983 | 0.081 | 0.016 | 0.073 | 0.045 | −0.152 | 0.23 | ⋯ | Prasad et al. (2014) |
| USPEB7 | 0.1923 | 0.86 | Pp | 3.40 | 3991 | 3966 | −0.390 | −0.456 | −0.333 | −0.347 | −1.045 | ⋯ | ⋯ | Priya et al. (2020) |
| BGVul | 0.4032 | 0.88 | Pp | 45.00 | 5868 | 5520 | 0.083 | 0.028 | 0.075 | 0.051 | −0.139 | 13.20 | ⋯ | Tanrıver (2014) |
| DZ Lyn | 0.3780 | 0.89 | Pp | 18.00 | 6860 | 5068 | 0.063 | 0.011 | 0.048 | 0.026 | −0.180 | ⋯ | ⋯ | Massimiliano et al. (2020) |
| LP Uma | 0.3099 | 0.89 | Pp | 14.00 | 5794 | 4921 | 0.000 | −0.052 | −0.032 | −0.054 | −0.314 | 12.50 | ⋯ | Prasad et al. (2014) |
| | 0.2286 | 0.90 | Pp | ⋯ | 5850 | 5758 | 0.055 | −0.002 | 0.090 | 0.080 | −0.270 | ⋯ | ⋯ | Elkhateeb et al. (2014) |

**Table 8**
(Continued)

| Variable | $P$ (days) | $q$ | Method[a] | $f\%$ | $T_h$ (K) | $T_c$ (K) | $\log M_h$ ($M_\odot$) | $\log M_c$ ($M_\odot$) | $\log R_h$ ($R_\odot$) | $\log R_c$ ($R_\odot$) | $\log J_{rel}$ | $\dot{P}$ ($10^{-7}$ day yr$^{-1}$) | $M_3$ ($M_\odot$) | Reference |
|---|---|---|---|---|---|---|---|---|---|---|---|---|---|---|
| 1SWASP J210318.76 +021002.2 | | | | | | | | | | | | | | |
| AV Pup | 0.4350 | 0.90 | Pp | 10.90 | 6255 | 6145 | 0.106 | 0.061 | 0.104 | 0.085 | −0.082 | 4.80 | ⋯ | Han et al. (2019) |
| OGLE J004619.65-725056.2 | 0.3766 | 0.91 | Pp | 34.63 | 5850 | 5734 | 0.061 | 0.020 | 0.045 | 0.028 | −0.175 | ⋯ | ⋯ | Priya & Rukmini (2018) |
| USPEB4 | 0.2007 | 0.92 | Pp | 3.00 | 4005 | 3972 | −0.354 | −0.389 | −0.256 | −0.277 | −0.952 | ⋯ | ⋯ | Priya et al. (2020) |
| KIC 7950962 | 0.8270 | 1.02 | Pp | 6.60 | 6800 | 6599 | 0.297 | 0.306 | 0.355 | 0.358 | 0.387 | ⋯ | ⋯ | Li & Liu (2020) |
| FV CVn | 0.3154 | 1.07 | Pp | 4.60 | 5470 | 5120 | −0.025 | 0.004 | −0.040 | −0.028 | −0.244 | −11.00 | ⋯ | Michel et al. (2019) |
| USPEB5 | 0.2040 | 1.09 | Pp | 2.90 | 3706 | 3622 | −0.334 | −0.373 | −0.292 | −0.310 | −0.920 | ⋯ | ⋯ | Priya et al. (2020) |
| USPEB2 | 0.2022 | 1.13 | Pp | 6.40 | 3597 | 3565 | −0.337 | −0.390 | −0.265 | −0.333 | −0.938 | ⋯ | ⋯ | Priya et al. (2020) |
| FP Lyn | 0.3591 | 1.15 | Pp | 13.00 | 5909 | 5255 | −0.015 | 0.047 | 0.003 | 0.029 | −0.134 | 4.19 | ⋯ | Michel et al. (2019) |
| V53 | 0.3085 | 1.23 | Pp | 79.70 | 7350 | 7523 | −0.088 | 0.002 | −0.067 | −0.028 | −0.217 | ⋯ | ⋯ | Liu et al. (2011) |
| V479 Lac | 0.3458 | 1.26 | Pp | 2.60 | 5713 | 5652 | −0.061 | 0.038 | −0.023 | 0.019 | −0.134 | −16.90 | ⋯ | Kjurkchieva et al. (2019) |
| 2MASS J20034981 +4411085 | 0.3667 | 1.29 | Pp | 21.00 | 5173 | 5160 | −0.053 | 0.058 | −0.003 | 0.044 | −0.087 | ⋯ | ⋯ | Joshi et al. (2016) |
| WTS 07g-3-00820 | 0.2270 | 1.30 | Pp | 5.00 | 5100 | 4967 | −0.191 | −0.305 | −0.148 | −0.034 | −0.730 | ⋯ | ⋯ | Gao et al. (2017) |
| V2252 Cyg | 0.2785 | 1.34 | Pp | 16.89 | 5590 | 5295 | −0.154 | −0.028 | −0.118 | −0.065 | −0.260 | ⋯ | ⋯ | Bulut et al. (2018) |
| V523 Aur | 0.3304 | 1.34 | Pp | 21.00 | 5500 | 5125 | −0.101 | 0.026 | −0.050 | 0.005 | −0.145 | ⋯ | ⋯ | Kjurkchieva et al. (2021) |
| FG Sct | 0.2706 | 1.35 | Pp | 21.40 | 4536 | 4373 | −0.167 | −0.037 | −0.131 | −0.076 | −0.277 | −0.64 | ⋯ | Yue et al. (2019) |
| OGLE J003835.24-735413.2 | 0.2691 | 1.37 | Pp | 19.57 | 4725 | 4509 | −0.167 | −0.055 | 0.060 | −0.052 | −0.476 | ⋯ | ⋯ | Priya & Rukmini (2018) |
| V336 TrA | 0.2668 | 1.40 | Pp | 15.70 | 5000 | 4840 | −0.046 | −0.192 | −0.076 | 0.070 | −0.492 | ⋯ | ⋯ | Kriwattanawong et al. (2018) |
| V2840 Cygni | 0.6311 | 1.36 | Pp | 12.39 | 6657 | 5652 | 0.103 | 0.230 | 0.212 | 0.266 | 0.288 | −5.48 | 0.51 | Current Work |

**Note.**
[a] S is the mass ratio of an eclipsing binary determined using spectroscopy; Tp is the mass ratio of a totally eclipsing binary determined using photometry and Pp is the mass ratio of a partially eclipsing binary determined using photometry.

relatively lower differences. Among these systems, some of the primary components exhibited stronger Hα lines than secondaries such as V2840 Cygni. In the case of V2840 Cygni, though the secondary component has a relatively deeper convective zone than the primary component, the ongoing mass transfer under the subsurface might have resulted in deeper absorption lines than the primary.

Seventeen of the cataloged HMRCBs including V2840 Cygni were also observed in LAMOST (Section 5). The spectral signatures of magnetic activity can be the contribution of one or both the components, since all the cataloged HMRCB components are of later spectral types. The majority of the HMRCBs are found to exhibit various characteristic lines such as Balmer lines, Na doublet, Mg triplet, Ca II H & K and Ca II IRT in the LAMOST spectra. Additional information on significant characteristic features of HMRCBs from the available literature is presented in Table 6.

HMRCBs including V2840 Cygni exhibited magnetic activity, but at various levels, which can be due to their magnetic activity cycle (quiet/active) or evolutionary state (mass transfer), or both. In the case of V2840 Cygni, there was no evidence of the magnetic activity cycle in the period variation study. In a spectroscopic study, the observed absorption profiles can be due to the subsurface hot spot region on the secondary formed by the rapid mass transfer from the primary component. The surface magnetic activity (photosphere) and subsurface hot spot region may explain the strong core fill-in absorption profiles observed at egress of the secondary eclipse. From the available LAMOST data of 17 HMRCBs, the spectral line analysis, including precise strengths of characteristic lines, is planned for future publication, which may lead to better understanding of magnetic activity in these systems.

The evolution of CBs on the basis of AML and TRO models is strongly examined only through observational studies. Following the TRO model, HMRCBs have been shown to be an important evolutionary phase. V2840 Cygni, being in the pivotal evolutionary phase, stands out in characterizing HMRCBs. The results obtained in the present work suggest that V2840 Cygni is a CB exhibiting a weak-contact configuration with a decreasing orbital period, and a subsurface hotspot region (evident through absorption profiles in observed spectra) due to mass transfer (in good agreement with the TRO model for observed HMR). Moreover, the system is evolving into a deeper-contact configuration. Construction of a robust catalog of HMRCBs with greater sample size leading to characteristic studies may aid in validating the relationship between the theoretical models and observations. Hence, additional high-resolution spectroscopic and long-term photometric observations in the future may help in affirming the models studied.

## Acknowledgments

We acknowledge

1. The Director of the Indian Institute of Astrophysics (IIA), Bengaluru for providing us with observation time at HCT and the supporting technical staff for helping with observations in service mode.
2. The use of TESS High Level Science Products (HLSP) produced by the Quick-Look Pipeline (QLP) at the TESS Science Office at MIT, which are publicly available from the Mikulski Archive for Space Telescopes (MAST). Funding for the TESS mission is provided by NASA's Science Mission directorate.
3. The use of spectra which are made publicly available through LAMOST DR7. The Guoshoujing Telescope (the Large sky Area Multi-Object Fiber Spectroscopic Telescope, LAMOST) for the spectroscopic data is a National Major Scientific Project built by the Chinese Academy of Sciences. Funding for the project has been provided by the National Development and Reform Commission. LAMOST is operated and managed by the National Astronomical Observatories, Chinese Academy of Sciences.
4. The use of data from the first public release of the WASP data (Butters et al. 2010) as provided by the WASP consortium and services at the NASA Exoplanet Archive, which is operated by the California Institute of Technology, under contract with the National Aeronautics and Space Administration under the Exoplanet Exploration Program.
5. The use of data from the European Space Agency (ESA) mission Gaia (https://www.cosmos.esa.int/Gaia), processed by the Gaia Data Processing and Analysis Consortium (DPAC, https://www.cosmos.esa.int/web/Gaia/dpac/consortium). Funding for the DPAC has been provided by national institutions, in particular the institutions participating in the Gaia Multilateral Agreement.
6. The use of data from the All-Sky Automated Survey for Supernovae (ASAS-SN) database.
7. The online resources which were used in the present work: the SIMBAD database, the ViZieR Catalog Service operated by CDS, Strasbourg, France, NASA ADS (Astrophysics Data System) and Google Scholar.